%% file: J1011.tex
\documentclass{emulateapj}
\usepackage{graphicx}
\usepackage{mathrsfs}	
\usepackage{amsmath}
\usepackage{color}
\usepackage{multirow}

\shorttitle{Mass/Light Offsets in the Einstein Cross---SDSS\,J1011$+$0143}
\shortauthors{Shu et al. 2015}
 
\begin{document}
 

\title{Kiloparsec Mass/Light Offsets in the Galaxy Pair-L\MakeLowercase{y}$\alpha$ Emitter Lens System SDSS\,J1011$+$0143$^{\dag}$}

\altaffiltext{$^{\dag}$}{Based on observations made with the NASA/ESA Hubble Space Telescope, obtained from the Data Archive at the Space Telescope Science Institute, which is operated by AURA, Inc., under NASA contract NAS 5-26555. These observations are associated with program \#10831.}

\author{\mbox{Yiping Shu\altaffilmark{1,2}}}
\author{\mbox{Adam S. Bolton\altaffilmark{2}}}
\altaffiltext{1}{National Astronomical Observatories, Chinese Academy of Sciences, 20A Datun Road, Chaoyang District, Beijing 100012, China ({\tt yiping.shu@nao.cas.cn})}
\altaffiltext{2}{Department of Physics and Astronomy, University of Utah,
115 South 1400 East, Salt Lake City, UT 84112, USA}

\author{\mbox{Leonidas A. Moustakas\altaffilmark{3}}}
\altaffiltext{3}{
Jet Propulsion Laboratory, California Institute of Technology, MS 169-506, 4800 Oak Grove Drive, Pasadena, CA 91109, USA
}
\author{\mbox{Daniel Stern\altaffilmark{3}}}


\author{\mbox{Arjun Dey\altaffilmark{4}}}
\altaffiltext{4}{National Optical Astronomy Observatory, Tucson, AZ 85719, USA
}

\author{\mbox{Joel R. Brownstein\altaffilmark{2}}}

\author{\mbox{Scott Burles\altaffilmark{5}}}
\altaffiltext{5}{Cutler Group, LP, 101 Montgomery Street, Suite 700, San Francisco, CA 94104, USA
}

\author{\mbox{Hyron Spinrad\altaffilmark{6}}}
\altaffiltext{6}{Department of Astronomy, University of California at Berkeley, Berkeley, CA 94720, USA
}

\begin{abstract}

We report the discovery of significant mass/light offsets in the strong gravitational lensing system SDSS\,J1011$+$0143. We use the high-resolution \textsl{Hubble Space Telescope} (\textsl{HST}) F555W- and F814W-band imaging and Sloan Digital Sky Survey (SDSS) spectroscopy of this system, which consists of a close galaxy pair with a projected separation of $\approx 4.2$ kpc at $z_{\rm lens} \sim 0.331$ lensing a Ly$\alpha$ emitter (LAE) at $z_{\rm source} = 2.701$. Comparisons between the mass peaks inferred from lens models and light peaks from \textsl{HST} imaging data reveal significant spatial mass/light offsets as large as $1.72 \pm 0.24 \pm 0.34$ kpc in both filter bands. Such large mass/light offsets, not seen in isolated field lens galaxies and relaxed galaxy groups, may be related to the interactions between the two lens galaxies. The detected mass/light offsets can potentially serve as an important test for the self-interacting dark matter model. However, other mechanisms such as dynamical friction on spatially differently distributed dark matter and stars could produce similar offsets. Detailed hydrodynamical simulations of galaxy-galaxy interactions with self-interacting dark matter could accurately quantify the effects of different mechanisms. The background LAE is found to contain three distinct star-forming knots with characteristic sizes from 116 pc to 438 pc. It highlights the power of strong gravitational lensing in probing the otherwise too faint and unresolved structures of distance objects below subkiloparsec or even 100 pc scales through its magnification effect.
\end{abstract}

\keywords{dark matter---galaxies: individual (SDSS\,J1011$+$0143)---galaxies: interactions---gravitational lensing: strong---techniques: image processing}

\slugcomment{Submitted to the ApJ}

\maketitle

\section{Introduction}
\label{sect_intro}
\input{sect_intro}

\section{The Data}
\label{sect_data}
\input{sect_data}

\section{Lens Modeling Strategy}
\label{sect_model}
\input{sect_model}

\section{Results}
\label{sect_results}
\input{sect_results}

\section{Discussion}
\label{sect_discussion}
\input{sect_discussion}

\section{Conclusion}
\label{sect_conclusion}
\input{sect_conclusion}

\acknowledgments

We thank the anonymous referee for helpful comments. 
The support and resources from the Center for High Performance Computing at the University of Utah are gratefully acknowledged. 
The work of L.A.M. and D.S. was carried out at Jet Propulsion Laboratory, California Institute of Technology, under a contract with NASA. 
A.D.'s research was supported by the National Optical Astronomy Observatory (NOAO). NOAO is operated by the Association of Universities for Research in Astronomy (AURA), Inc., under a cooperative agreement with the National Science Foundation. 
Support for program \# 10831 was provided by NASA through a grant from the Space Telescope Science Institute, which is operated by the Association of Universities for Research in Astronomy, Inc., under NASA contract NAS 5-26555. 

Funding for the SDSS and SDSS-II has been provided by the Alfred P. Sloan Foundation, the Participating Institutions, the National Science Foundation, the U.S. Department of Energy, the National Aeronautics and Space Administration, the Japanese Monbukagakusho, the Max Planck Society, and the Higher Education Funding Council for England. The SDSS Web site is http://www.sdss.org/.

The SDSS is managed by the Astrophysical Research Consortium for the Participating Institutions. The Participating Institutions are the American Museum of Natural History, Astrophysical Institute Potsdam, University of Basel, University of Cambridge, Case Western Reserve University, University of Chicago, Drexel University, Fermilab, the Institute for Advanced Study, the Japan Participation Group, Johns Hopkins University, the Joint Institute for Nuclear Astrophysics, the Kavli Institute for Particle Astrophysics and Cosmology, the Korean Scientist Group, the Chinese Academy of Sciences (LAMOST), Los Alamos National Laboratory, the Max-Planck-Institute for Astronomy (MPIA), the Max-Planck-Institute for Astrophysics (MPA), New Mexico State University, Ohio State University, University of Pittsburgh, University of Portsmouth, Princeton University, the United States Naval Observatory, and the University of Washington.


\end{document}

%% file: sect_intro.tex
Galaxies are the fundamental building blocks of the visible Universe. 
They contain all the known ingredients in the Universe including stars, 
gas, and dark matter.
In the currently well-established $\Lambda$ cold dark matter ($\Lambda$CDM) 
scenario \citep[e.g.,][]{Riess98, Perlmutter99, Burles01, Anderson14, Planck15}, 
galaxies form in the so-called bottom-up hierarchical 
clustering fashion, with low-mass galaxies and dark matter halos 
that galaxies reside in forming early and subsequently growing and merging 
to form more massive galaxies/halos 
\citep[e.g.,][]{Peebles68, White78, Blumenthal84, White91, Kauffmann93, Navarro95}. 

Interacting galaxies serve as valuable astronomical laboratories, allowing 
the study of different species of particles under violent dynamical conditions, 
and hence augmenting our understanding of the natures of stars, gas, and dark matter. 
By monitoring X-ray and H\textsc{i} data in galaxy clusters and merging galaxies, 
gas is found to fall behind stars and dark matter 
during galaxy interactions \citep[e.g.,][]{Cayatte90, Allen02, Markevitch02, 
Clowe04, Kenney04, Markevitch04, Bradavc06, Randall08, Million10, Kohlinger14}. 
This lagging is consistent with the ram pressure stripping picture 
first proposed by \citet{Gunn72}, in which the interstellar gas is slowed down 
by the ram pressure of the intracluster medium. 
Stars are barely affected as they are much more massive and tightly bound. 

Owing to its dark nature, 
dark matter can only be probed by a limited number of techniques. 
Being extremely sensitive to gravity, the strong gravitational lensing effect is 
the most powerful and promising probe of dark matter.  
Utilizing multiple images seen in strong gravitational lensing systems, 
people have confirmed the existence of dark matter and explored its properties 
to a great extent \citep[e.g.,][]{Kochanek95, Keeton98, Rusin03, 
SLACSV, SLACSVII, Vegetti12, Brownstein12, 
Bolton12, Nierenberg14, Shu15}. 
Recently, several studies have detected a further separation between 
the distributions of stars and total mass in galaxy clusters 
\citep[e.g.,][]{Williams11, Mohammed14, Harvey15, Massey15}. 
Such separations can be used to test the self-interacting dark matter model, 
because dark matter self-interactions would induce an extra drag force 
only on dark matter and therefore separate it from stars. 

Indeed, several recent works have estimated the lower limits on the dark matter 
self-interacting cross section assuming that the detected 
mass/light offsets are solely caused by dark matter self-interactions 
\citep{Williams11, Kahlhoefer15, Massey15}. 
However, other possibilities need to be taken into account as well 
when interpreting such separations 
\citep[e.g.,][]{Kahlhoefer14, Mohammed14, Massey15}. 

On the other hand, significant mass/light offsets have never been detected in 
galaxy-scale or galaxy-group-scale systems. 
Extensive studies of over 200 isolated strong lens galaxies 
discovered by the Lenses Structure and Dynamics (LSD) Survey, 
the Sloan Lens ACS (SLACS) Survey, 
the BOSS Emission-Line Lens Survey (BELLS), 
the Strong Lensing in the Legacy Survey (SL2S), 
and the SLACS for the Masses (S4TM) Survey 
have verified that in general light traces mass well in terms of 
the coincidence of centroids 
\citep[e.g.,][]{Treu04, SLACSIII, SLACSIV, SLACSV, Brownstein12, 
Sonnenfeld13, Newman15, Shu15}. 
To date, approximately 100 group-scale strong lenses have been identified in the 
Cambridge And Sloan Survey Of Wide ARcs in the skY (CASSOWARY) Survey 
\citep{Stark13}, the Sloan Bright Arcs Survey \citep{Kubo10}, 
and the Strong Lensing Legacy Survey-ARCS (SARCS) sample \citep{More12}, 
as well as other individual observations. 
No mass/light offsets have been reported in these group-scale lenses either, 
although some works explicitly fix the mass centroid to that of the light 
for systems with insufficient constraints 
\citep[e.g.,][]{Dye08, Belokurov08, Limousin09, Suyu09, Jones10, Grillo11, 
Oguri12, Grillo13, Grillo14, Foex14}. 
Nevertheless, the nondetections suggest that these isolated lens galaxies and 
galaxy groups are substantially relaxed. 

In this paper, we present the first detection of significant mass/light offsets 
on galaxy scales. 
The system we study is the strong gravitational lens system SDSS\,J1011$+$0143, 
which was first discovered by \citet{Bolton06b} and further followed up with 
high-resolution \textsl{Hubble Space Telescope} (\textsl{HST}) imaging observations. 
A galaxy pair consisting of two early-type galaxies (ETGs) with a projected distance 
of $0.88{\arcsec}$ at redshifts of $\sim 0.331$ acts as the lens, 
while the background source is an Ly$\alpha$ emitter (LAE) at $z_{\rm source}=2.701$. 
We perform lens modeling based on the \textsl{HST} 
F555W- and F814W-band imaging data and construct the total mass distributions 
in both filter bands for SDSS\,J1011$+$0143. 
The mass-model results from both filter bands are in excellent agreement. 
The mass peaks inferred from lens models exhibit clear spatial offsets 
up to $0\farcs36 \pm 0\farcs05 \pm 0\farcs07$, 
or equivalently $1.72 \pm 0.24 \pm 0.34$ kpc, 
from the star-light peaks seen in both filter bands. 
We suggest that the offsets in SDSS\,J1011$+$0143 
are caused by the interactions of the lensing galaxy pair. 

The paper is organized as follows. 
Section~\ref{sect_data} introduces the \textsl{HST} imaging data used. 
Section~\ref{sect_model} explains our lens modeling strategy. 
Results are summarized in Section~\ref{sect_results}. 
We discuss the detected mass/light offsets in Section~\ref{sect_discussion}, 
and we present the conclusions in Section~\ref{sect_conclusion}.
For all calculations, we adopt a fiducial cosmological model with $\rm \Omega_m = 0.274$, $\rm \Omega_{\Lambda} = 0.726$ and $H_0 \rm = 70\,km\,s^{-1}\,Mpc^{-1}$ \citep[WMAP7;][]{WMAP7}.

%% file: sect_data.tex
SDSS\,J1011$+$0143 was first identified as a strong gravitational lens system 
by \citet{Bolton06b} based on the presence of an anomalous high-significance 
emission line at $\lambda_{\rm obs} \approx 4500$\,\AA\ in its 
Sloan Digital Sky Survey (SDSS) spectrum. 
Follow-up long-slit spectroscopic and imaging observations 
using the Low Resolution Imaging Spectrometer \citep[LRIS;][]{Oke95} 
on the Keck I telescope confirmed the spectroscopic line detection 
and revealed quadruple lensed images in a symmetric configuration. 
The foreground lens, appearing to be a single component in both the SDSS 
and \citet{Bolton06b} observations, had a spectroscopic redshift of 
$z_{\rm lens}=0.331$. 
The anomalous emission line, coincident with the lensed images on either side of the 
lens galaxy as indicated by the LRIS long-slit spectroscopy, showed the classic, 
asymmetric morphology typical of Ly$\alpha$ emission 
($\lambda_{\rm rest} = 1215.67$\,\AA). 
It was therefore determined to be a Ly$\alpha$ emission coming from a background 
LAE at $z_{\rm source} = 2.701$ because there are no other lines detected such 
as would be expected if it were something else. 
Based on the LRIS $B$-band imaging data, \citet{Bolton06b} found the Einstein radius 
of SDSS\,J1011+0143 to be $\theta_{\rm E} \simeq 1\farcs84$ 
and the total mass within $\theta_{\rm E}$ to be 
$(5.2 \pm 0.1) \times 10^{11} M_{\odot}$.

The data that we use in this work were collected by the follow-up \textsl{HST} GO Program\# 10831 (PI L. Moustakas) through the F555W (V) and F814W (I) filters of the Wide-Field Channel (WFC) of the Advanced Camera for Surveys (ACS)\@. For each filter, 4 subexposures, each with an exposure time of 522 seconds, were successfully taken in 2006 November and December. The archival flat-fielded (FLT) subexposure data for each filter were fully reduced and combined using the custom-built GUI tool, \emph{ACSPROC}, following the recipe described in \citet{SLACSV}, \citet{Brownstein12}, and \citet{Shu15}. 

\begin{figure}[htbp]
\centering
\includegraphics[width=0.45\textwidth]{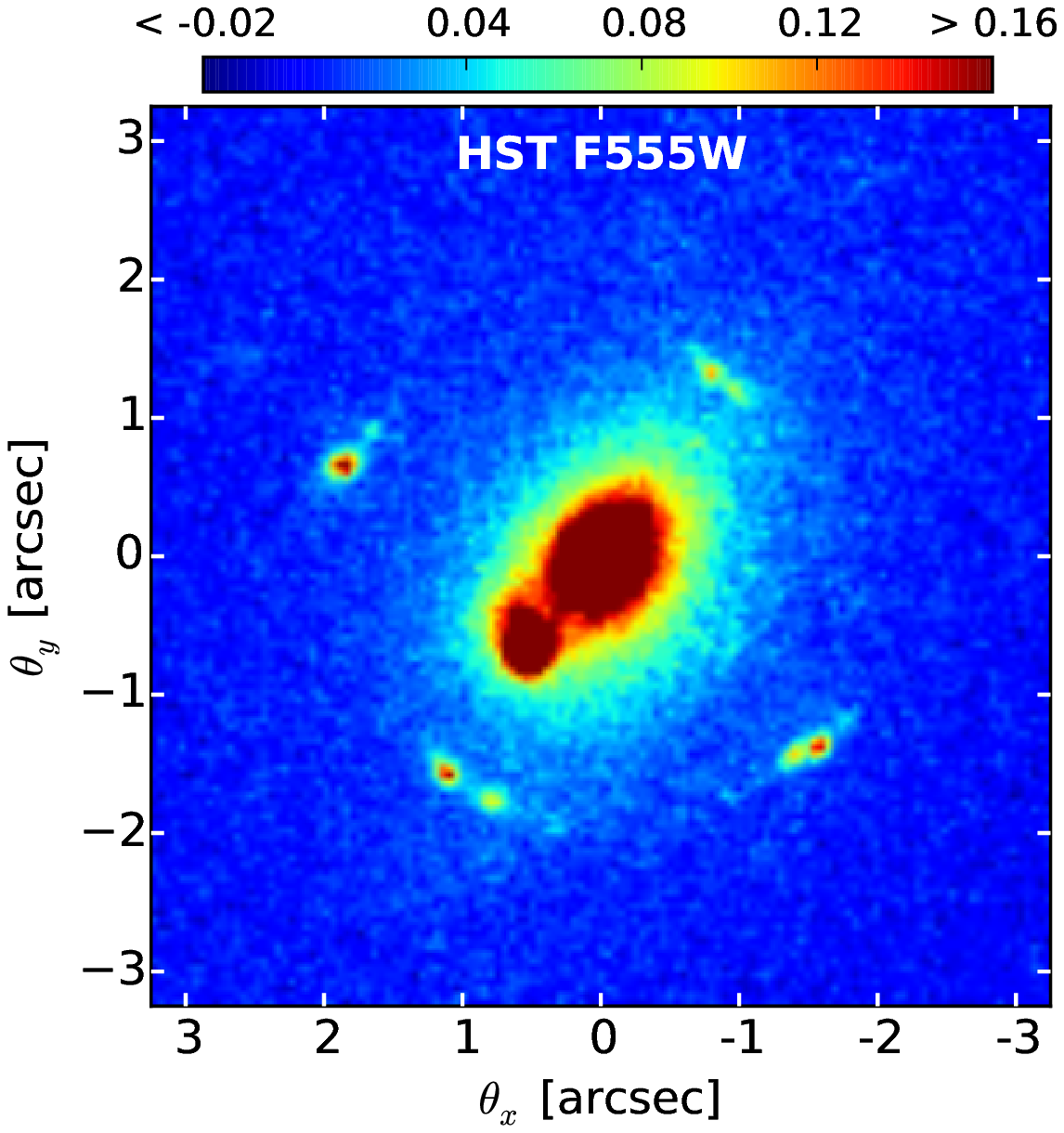}
\includegraphics[width=0.45\textwidth]{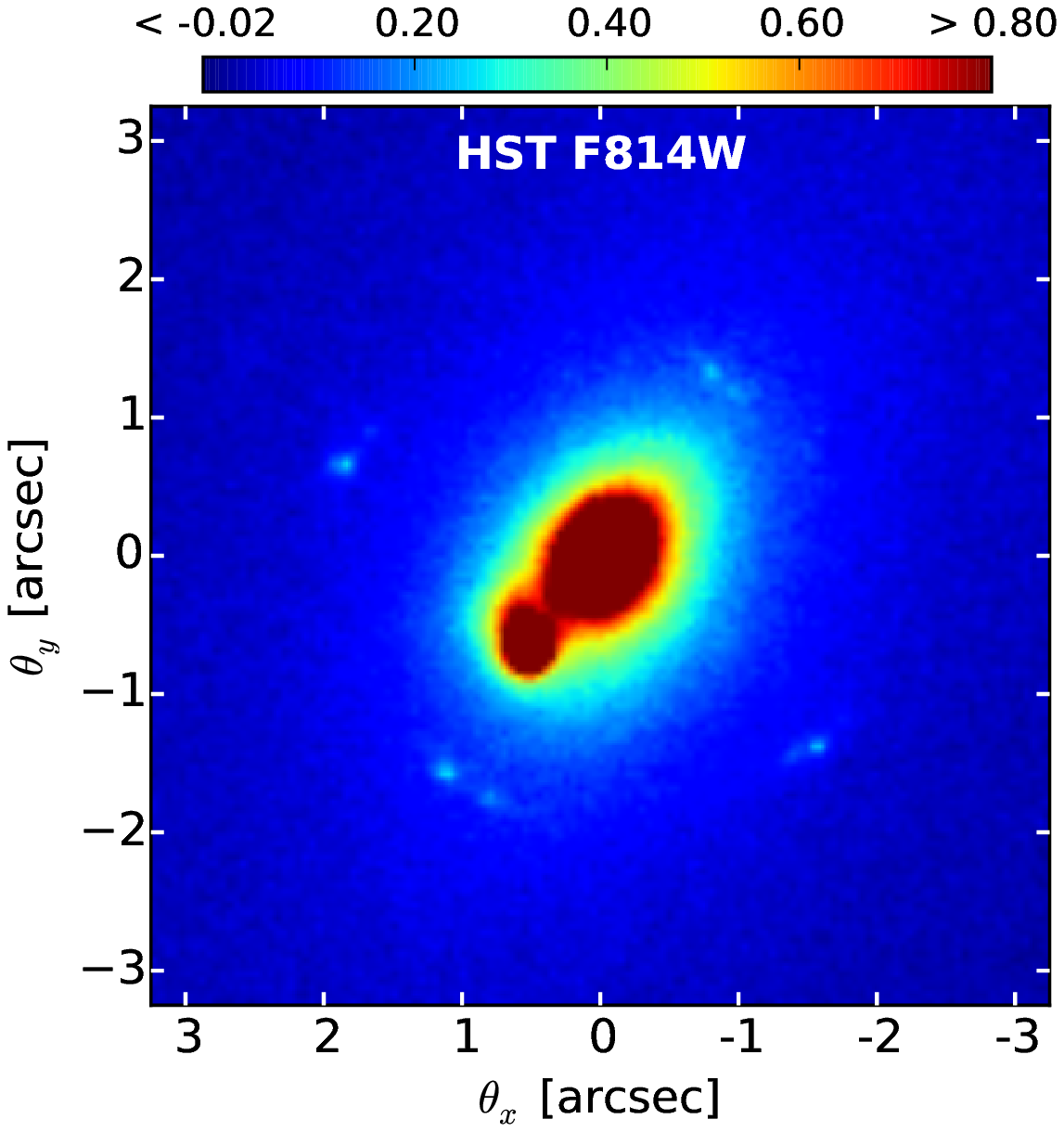}
\caption{
\label{fig:mosaic}
Central $6\farcs5 \times 6\farcs5$ mosaics of the lens system SDSS\,J1011$+$0143 viewed in the \textsl{HST} F555W (top) and F814W (bottom) bands. All images displayed in this paper are oriented such that north is up and east is to the left, with units of $x$ and $y$ giving offsets in R.A. (J2000) and decl. (J2000) relative to the center ($10^{\rm h} 11^{\rm m} 29\fs49$, $+01\degr 43\arcmin 23\farcs25$). The color bars indicate the intensity levels in units of electrons per second per pixel$^2$.}
\end{figure}

Figure~\ref{fig:mosaic} shows the reduced central $6\farcs5 \times 6\farcs5$ mosaics of the lens system SDSS\,J1011$+$0143 viewed in the \textsl{HST} F555W band (top) and F814W band (bottom). 
The overall configuration of the system looks alike in both filter bands. 
The high spatial resolution of the \textsl{HST} reveals that the foreground lens actually comprises two distinct luminous components. The primary lens galaxy at which the mosaics are centered is brighter and more extended, while the secondary lens is located $\sim 0\farcs88$ away to the southeast. 
The spatially resolved four lensed images also indicate a multicomponent source configuration. 
Although no spectroscopic data are currently available for each individual lens galaxy, neither the central $1\farcs5$-radius aperture-integrated spectrum from SDSS nor the long-slit observation going across the minor axis of the unresolved lens shows any noticeable structures in the absorption lines indicating two different redshifts. As will be shown later, the two lens galaxies also have the same {\it V-I} color (within the errors). Therefore, they must have similar redshifts. That is further supported by the slightly distorted morphological shapes of the two lens galaxies and a possible ``bridge'' between them as indications of recent/ongoing galaxy interactions. Considering that the SDSS-measured velocity dispersion of the unresolved system is $259 \pm 16$ km\,s$^{-1}$ with a resolution of 70 km\,s$^{-1}$, we estimate the relative line-of-sight velocity of the two lens galaxies to be $\lesssim 350$ km\,s$^{-1}$.

The major difference between the F555W band and F814W band is in the contrast between the foreground lens galaxies and lensed images. 
The lens galaxies are much brighter ($\sim 6\times$) viewed in the F814W band, while 
the brightnesses of the lensed images stay about the same in the two bands. 
As a result, the image contrast is much more favorable for the study of the relatively faint lensed source in the F555W band. This is exactly as expected because the background source is an LAE, presumably with active star formation, and should appear bluer in color as compared to the foreground lenses that are red ETGs. 

We note here that for all the \textsl{HST} imaging data, the corresponding pixel count errors are rescaled such that the average error of the background matches the standard deviation of the background to correct for possible correlations in the errors caused by the image resampling procedure in the data reduction.

%% file: sect_model.tex
In order to obtain an accurate lens model from the two-dimensional imaging data, 
the surface brightness distribution of the foreground lens needs to be 
appropriately modeled, especially when it contributes a significant fraction 
to the light budget at positions of the lensed features. 
As shown by Figure~\ref{fig:mosaic}, SDSS\,J1011$+$0143 consists of two luminous lens 
galaxies with distorted morphologies. A simple two-component fit to the foreground 
light assuming a S{\'e}rsic model \citep{Sersic63} 
leaves two relatively localized clumps of residuals around the centers of 
the two lens galaxies. 
We therefore build a four-S{\'e}rsic component model for the foreground-light 
distribution.

To generate the lens model, we employ the parametric lens fitting technique 
implemented in \citet{Shu15} with appropriate modifications. 
The total surface mass distribution is assumed to be contributed solely from 
the lens galaxy pair characterized by two singular isothermal ellipsoid (SIE) 
mass clumps. The total deflection angle is calculated analytically following 
\citet{Kormann94}. We continue adding S{\'e}rsic blobs one by one to 
model the surface brightness distribution of the background source 
until a satisfactory fit is achieved. In the end the source is modeled by 
three S{\'e}rsic blobs. 
The predicted lensed images are generated by forward ray tracing. 

\begin{figure*}[htbp]
\centering
\includegraphics[width=0.99\textwidth]{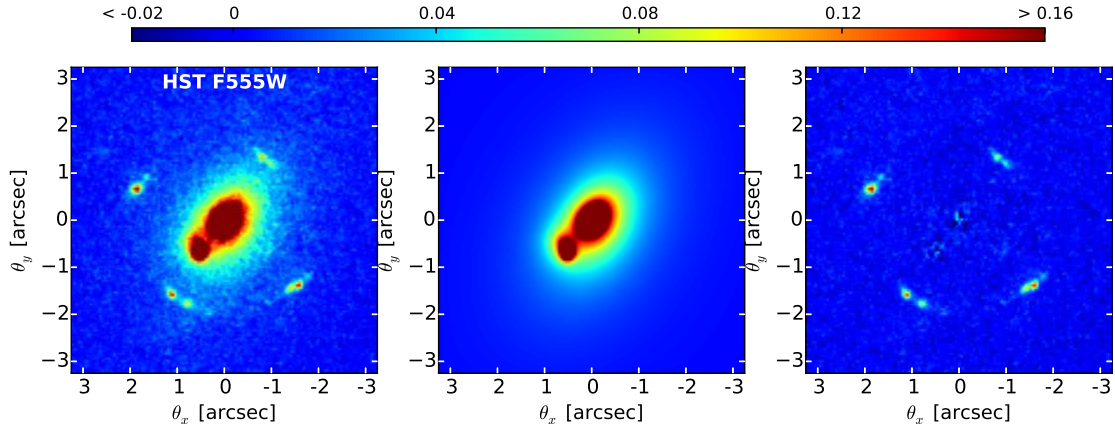}
\caption{
\label{fig:foreground_subtraction_F555W}
Performance of the foreground-light subtraction with the four S{\'e}rsic model 
in the \textsl{HST} F555W band. 
The reduced data in the F555W band, the best-fit model, and 
the foreground-subtracted residual are shown from left to right, respectively. 
}
\end{figure*}
\input{tb_foreground_parameters}

Instead of performing lens modeling on the foregroud-light-subtracted data as 
is commonly done in the community 
\citep[e.g.,][]{SLACSIII, Marshall07, SLACSV, SLACSXI, Brewer12, Brownstein12, 
Sonnenfeld13, Shu15}, 
we fit the foreground light and mass simultaneously in this work. 
This treatment has the advantage of largely reducing the systematic errors 
in the inferred model parameters that, 
as discussed by \citet{SLACSI} and \citet{Marshall07},
get introduced by adopting different 
foreground-light subtraction methods with different parameterized models, 
especially when the foreground-light subtraction and lens modeling are 
done separately. 
The foreground-light distribution model is combined with the predicted 
lensed images and convolved with the appropriate point spread function (PSF) 
generated by the {\tt Tiny Tim} tool \citep{Krist93}. 
The combined model is then compared to the observed data, 
and the goodness of fit is determined by the $\chi^2$ function, defined as
\begin{equation}
\chi^2 = \sum_{i, j} [\frac{I^{\rm data}_{i, j}-(I^{\rm lens}_{i, j}+I^{\rm image}_{i, j})}{\sigma_{i, j}}]^2,
\label{eq:chi2}
\end{equation}
where $I^{\rm data}_{i, j}$, $I^{\rm lens}_{i, j}$, 
and $I^{\rm image}_{i, j}$ are the observed, PSF-convolved foreground lens, 
and PSF-convolved lensed image intensities at pixel $(i, j)$ in the image plane, 
respectively. $\sigma_{i, j}$ is the rescaled flux error at pixel $(i, j)$. 
All the model parameters are optimized using the Levenberg-Marquardt algorithm 
with the \emph{LMFIT} package \citep{lmfit}. 

We apply the aforementioned lens modeling strategy to the data in both filters 
independently. 
As will be shown later, the best-fit models based on the two different sets of data 
converge nicely within uncertainties. 
Since the F555W-band results should in principle be less affected by the 
foreground-light subtraction (owing to better contrast of the lensed images 
relative to the lens galaxies), we quote F555W lens model parameters for 
definiteness in the text and report the results obtained from 
the F814W-band data in the tables for reference.

\citet{Marshall07} explored the systematic uncertainty in the lens and source 
parameters introduced by different foreground-light subtraction schemes 
and found it to be $\sim 0.6\%$ in the Einstein radius, 
$\sim 2\%$ in the source effective radius, 
and $\sim 0.1$ mag in the source magnitude.
Several other studies quantified the overall systematic uncertainty in the 
Einstein radius to be $2-3 \%$ \citep{SLACSV, Sonnenfeld13}. 
Compared to those studies, 
our strategy of fitting the light and mass simultaneously should 
in principle reduce that part of the systematic uncertainties. 
Nevertheless, we adopt $3\%$ as a generous fractional uncertainty 
in the measured Einstein radius owing to foreground subtraction 
in subsequent analyses. As will be shown later, this $3\%$ uncertainty 
is only a minor effect when compared to the statistical uncertainty 
of the derived lens model parameters.

%% file: tb_foreground_parameters.tex
\begin{deluxetable*}{lcccccccc}
\tabletypesize{\scriptsize}
\tablewidth{\hsize}
\tablecaption{Foreground-light Model Parameters}
\tablehead{
\colhead{Band} &
\colhead{ID} &
\colhead{$m_{\rm AB} [mag]$} &
\colhead{$x_c[\arcsec]^*$} &
\colhead{$y_c[\arcsec]^*$} &
\colhead{$R_e[\arcsec]$} &
\colhead{$n$} &
\colhead{$\phi_*[\degr]$} &
\colhead{$q_*$}}
\startdata
\multirow{4}{*}{F555W} &
1.1 &
\multirow{2}{*}{$ 19.54 \pm  0.11$} &
$ -0.0095 \pm 0.0006$ &
\,\,\,\,$ 0.0254 \pm 0.0005$ &
$  0.217 \pm  0.008$ &
$   2.11 \pm   0.07$ &
$  141.8 \pm    0.6$ &
$  0.686 \pm  0.006$ \\ 
 &
1.2 &
 &
\,\,\,\,\,$  0.118 \pm  0.005$ &
$ -0.084 \pm  0.005$ &
$   1.28 \pm   0.04$ &
$   1.43 \pm   0.04$ &
$  150.6 \pm    0.5$ &
$  0.665 \pm  0.005$ \\ 
 &
2.1 &
\multirow{2}{*}{$ 22.14 \pm  0.11$} &
$ -0.5208 \pm 0.0005$ &
$ -0.6889 \pm 0.0008$ &
$ 0.0650 \pm 0.0014$ &
$   1.03 \pm   0.08$ &
$    169 \pm      3$ &
$  0.831 \pm  0.014$ \\ 
 &
2.2 &
 &
$ -0.607 \pm  0.006$ &
$ -0.496 \pm  0.009$ &
$  0.246 \pm  0.009$ &
$   0.93 \pm   0.06$ &
\,\,\,\,\,\,$     24 \pm     19$ &
$   0.95 \pm   0.04$ \\ 
\hline \\ [-1.8ex] 
\multirow{4}{*}{F814W} &
1.1 &
\multirow{2}{*}{$ 17.69 \pm  0.11$} &
$ -0.0085 \pm 0.0003$ &
\,\,\,\,$ 0.0260 \pm 0.0002$ &
$  0.274 \pm  0.006$ &
$   2.78 \pm   0.04$ &
$  143.1 \pm    0.3$ &
$  0.692 \pm  0.002$ \\ 
 &
1.2 &
 &
\,\,\,\,\,$  0.130 \pm  0.003$ &
$ -0.103 \pm  0.003$ &
$   1.70 \pm   0.03$ &
$   1.73 \pm   0.02$ &
$  149.8 \pm    0.2$ &
$  0.660 \pm  0.002$ \\ 
 &
2.1 &
\multirow{2}{*}{$ 20.38 \pm  0.11$} &
$ -0.5206 \pm 0.0002$ &
$ -0.6898 \pm 0.0004$ &
$ 0.0530 \pm 0.0006$ &
$   1.18 \pm   0.06$ &
$    169 \pm      1$ &
$  0.785 \pm  0.008$ \\ 
 &
2.2 &
 &
$ -0.601 \pm  0.002$ &
$ -0.513 \pm  0.004$ &
$  0.244 \pm  0.004$ &
$   1.05 \pm   0.03$ &
\,\,\,$     23 \pm      3$ &
\,\,$   0.87 \pm   0.02$ 
\tablecomments{Best-estimated parameter values of the foreground-light model. Columns (from left to right) represent the apparent magnitude, central $x$ and $y$ coordinates, effective radius, S{\'e}rsic index, position angle, and axis ratio of the individual S{\'e}rsic blob. \\
$*$: Positions relative to the center of the cutout mosaic with R.A. (J2000) and decl. (J2000) of ($10^{\rm h} 11^{\rm m} 29\fs49$, $+01\degr 43\arcmin 23\farcs25$).}
\enddata
\label{tb:foreground_parameters}
\end{deluxetable*}

%% file: sect_results.tex
\subsection{Best-fit Parameters}

\begin{figure*}[htbp]
\centering
\includegraphics[width=0.99\textwidth]{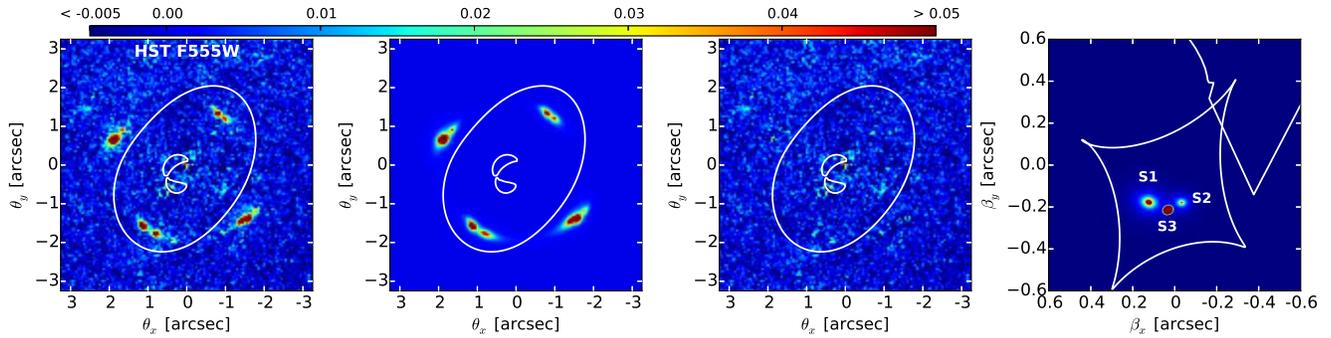}
\caption{
\label{fig:model}
Performance of the lens model in the F555W band. 
The foreground-subtracted data in the F555W band, 
the best-fit lens model, the final residual, and the source-light distribution 
are displayed from left to right, respectively. 
White lines in the image plane (first 3 columns) are the critical lines, while 
white lines in the source plane (last column) are the caustics. }
\end{figure*}
\input{tb_lens_parameters}
\input{tb_source_parameters}

Figure~\ref{fig:foreground_subtraction_F555W} illustrates the performance of 
the four-S{\'e}rsic model in the foreground-light subtraction. 
We obtain a satisfactory fit to the foreground-light distribution. 
As mentioned earlier, the performance is similarly good for the F814W-band data.
Table~\ref{tb:foreground_parameters} summarizes the best-estimated parameter 
values of the four-S{\'e}rsic model in both filter bands. 
We denote the four components as $1.1$, $1.2$, $2.1$, and $2.2$. 
We can see that both the primary lens and secondary lens can be 
well represented by a compact and bright component ($1.1$/$2.1$) and 
a more extended but relatively fainter component ($1.2$/$2.2$) 
with small separations ($\sim 0\farcs1-0\farcs2$) between the two. 
Combining light from 1.1 and 1.2, we estimate the {\it V-} and {\it I-}band AB 
apparent magnitudes for the primary lens to be 19.54 and 17.69, respectively. 
Similarly, the {\it V-} and {\it V-}band AB apparent magnitudes 
for the secondary lens are 22.14 and 20.38, respectively. 
For simplicity, we adopt the same error of 0.11 for the apparent 
magnitudes, or equivalently 10\% fractional error in the fluxes 
as suggested by the best-fit model. 
Note that the two lens galaxies have the same {\it V-I} color within the errors, 
being $1.85 \pm 0.16$ for the primary lens 
and $1.76 \pm 0.16$ for the secondary lens. 
The centroid position, position angle, and axis ratio of the four components 
are consistent within $3\sigma$ between the F555W and F814W bands. 
The effective radius and S{\'e}rsic index are different mainly because 
the two bands essentially trace different populations of stars 
that presumably distribute differently. 

Although the foreground-light distribution is complex, the best-fit 
lens mass model turns out to be rather simple. Figure~\ref{fig:model} 
demonstrates the performance of the two-SIE lens model in the F555W band. 
The first three panels display the foreground-subtracted F555W-band data, 
the best-fit model, and the final residual. 
Critical lines connecting points of infinite magnification in the image plane 
are also plotted in white. We can see from the residual plot that 
the two-SIE lens model is sufficient to provide a satisfactory recovery 
of the observed quadruple lensed images. 
The $\chi^2$ value of the fit defined by Equation~(\ref{eq:chi2}) is 17,032 for 
a degree of freedom (dof) of 17,101. 
The F814W-band data are also well explained by a set of similar lens model 
parameters with a $\chi^2$ value of 18,384 for the same dof.

The best-estimated lens model parameters inferred from individual filter bands 
are summarized in 
Table~\ref{tb:lens_parameters}. The first column is the lensing strength 
(Einstein radius) $b_{\rm SIE}$ in arcseconds. The reported uncertainty 
in $b_{\rm SIE}$ is a combination of the statistical uncertainty and 
the $3\%$ systematic uncertainty due to foreground subtraction in quadrature. 
The characteristic lensing velocity dispersion 
$\sigma_{\rm SIE}$ is related to $b_{\rm SIE}$ as
\begin{equation}
b_{\rm SIE} = 4 \pi \frac{\sigma_{\rm SIE}^2}{c^2} \frac{D_{\rm LS}}{D_{\rm S}}
\end{equation}
where $D_{\rm LS}$ and $D_{\rm S}$ are the angular diameter distances 
from the lens and the observer to the source, respectively. 
Here we adopt the intermediate-axis convention of \citet{Kormann94} 
for $b_{\rm SIE}$ and $\sigma_{\rm SIE}$. 
Focusing on the F555W band, the primary lens has a lensing velocity dispersion 
$\sigma_{\rm SIE}$ of $236\pm18$ km\,s$^{-1}$. The secondary lens has a 
slightly smaller lensing velocity dispersion of 
$\sigma_{\rm SIE}=169\pm25$ km\,s$^{-1}$. 

In the best-fit model, there are three distinct source components located 
within the caustics (the rightmost panel in Figure~\ref{fig:model}), 
which are the mappings of the critical lines in the source plane. 
We denote them as S1, S2, and S3 in descending order of brightness. 
Best-estimated values for a selection of source parameters are summarized in 
Table~\ref{tb:source_parameters}. 
The uncertainty in the source apparent magnitude is a combination of 
the statistical uncertainty and the $0.1$ mag systematic uncertainty due to 
foreground subtraction in quadrature. 
The statistical uncertainty in the total flux is assumed to be equal to that 
of the peak intensity, which ranges from $5\%$ to $\sim 50\%$. 
We define the average magnification $\mu$ for each source component to be 
the ratio of the total flux mapped onto the image plane to 
the total flux in the source plane. 
S1 is the most luminous component with an intrinsic {\it V-}band 
AB apparent magnitude of $26.7 \pm 0.6$. 
Its effective radius is $0\farcs054 \pm 0\farcs005$ in the source plane, 
which corresponds to a physical size of $438 \pm 38$ pc. 
S2 and S3 are relatively fainter and smaller with effective radii of 
$266 \pm 37$ pc and $116 \pm 6$ pc, respectively.
All three source components are highly magnified by a factor of 19 or more. 
Recalling that the background source is an LAE at a redshift of 2.701, 
the \textsl{HST} F555W filter covers the rest frame $\approx 1200-1600$ \AA. 
The detected three distinct source components hence 
correspond to individual star-forming knots at subkiloparsec scales in the LAE. 
It highlights the power of strong gravitational lensing in probing 
the otherwise too faint, unresolved ``fine'' structures of distance objects below 
subkiloparsec or even 100 pc scales through its magnification effect.

\subsection{Mass/Light Offsets}

\begin{figure}[htbp]
\centering
\includegraphics[width=0.45\textwidth]{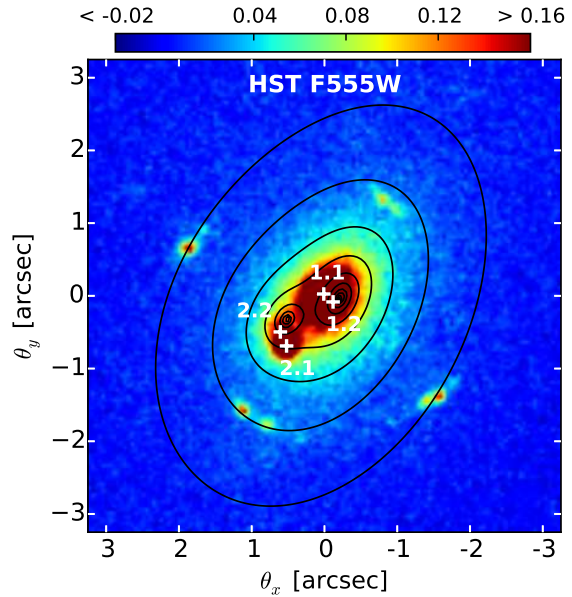}
\caption{
\label{fig:kappa}
Mass/light offsets in SDSS\,J1011$+$0143 in the \textsl{HST} F555W band. 
The black contours represent the surface mass isodensity levels. 
White plus signs mark the individual light peaks.}
\end{figure}

The centroids of the projected mass components exhibit significant spatial offsets 
from the centroids of the corresponding luminous components in the lens galaxies. 
The offsets are clearly visualized in Figure~\ref{fig:kappa}, where we overlay the 
projected total mass isodensity contours (black) reconstructed from the lens model 
on the \textsl{HST} F555W-band image. The foreground-light peaks are marked by the 
white plus signs. 
We associate the mass peak near light peak $1.2$ with the primary lens 
and the mass peak near light peak $2.2$ with the secondary lens. 
The projected spatial mass/light offsets for the primary lens are 
\begin{align}
\Delta_{1.1} &= 0\farcs23 \pm 0\farcs05 \pm 0\farcs07 = (1.08 \pm 0.23 \pm 0.34)\, \rm kpc \\
\Delta_{1.2} &= 0\farcs11 \pm 0\farcs05 \pm 0\farcs07 = (0.51 \pm 0.22 \pm 0.34)\, \rm kpc,
\end{align} 
where $\Delta_{1.1}$ and $\Delta_{1.2}$ are the offsets between 
the mass and light peaks $1.1$ and $1.2$, respectively. 
Similarly, for the secondary lens, 
\begin{align}
\Delta_{2.1} &= 0\farcs36 \pm 0\farcs05 \pm 0\farcs07 = (1.72 \pm 0.24 \pm 0.34)\, \rm kpc \\
\Delta_{2.2} &= 0\farcs19 \pm 0\farcs07 \pm 0\farcs07 = (0.92 \pm 0.36 \pm 0.34)\, \rm kpc. 
\end{align} 

Note that the first set of uncertainties in the above equations represent the 
statistical uncertainties from fitting the \textsl{HST} imaging data. The second 
set of uncertainties represent the systematic uncertainties in the data themselves 
including two major contributions, 
the $\sim 0\farcs05$ native pixel size of the \textsl{HST} WFC camera 
and the $\approx 0\farcs043$ half-width at half-maximum of the 
applied PSF. Nevertheless, the detected mass/light offsets are of high 
statistical significances, especially for $\Delta_{2.1}$. 

We point out that the offsets between the mass peak and 
the light peaks for the more massive primary lens 
($\Delta_{1.1}$ and $\Delta_{1.2}$) are relatively smaller than 
those for the less massive secondary lens ($\Delta_{2.1}$ and $\Delta_{2.2}$). 
In addition, $\Delta_{1.1}$ and $\Delta_{1.2}$ point, to the zeroth order, 
in the opposite direction of $\Delta_{2.1}$ and $\Delta_{2.2}$. 
The separations from the mass peaks to the fainter, more extended 
light components 1.2 and 2.2 are smaller than those to 
the brighter and more compact light components 1.1 and 2.1. 
The statistical significance of offsets $\Delta_{1.2}$ and $\Delta_{2.2}$ are 
$\sim 2.0\sigma$. The significance of $\Delta_{1.1}$ is approximately 2.7$\sigma$, 
and the significance of $\Delta_{2.1}$ is $\approx 4.2\sigma$.

%% file: tb_lens_parameters.tex
\begin{deluxetable*}{lccccccc}
\tabletypesize{\scriptsize}
\tablewidth{\hsize}
\tablecaption{Lens Model Parameters}
\tablehead{
\colhead{Band} &
\colhead{ID} &
\colhead{$b_{\rm SIE}[\arcsec]$} &
\colhead{$\sigma_{\rm SIE}[{\rm km\,s^{-1}}]$} &
\colhead{$x_c[\arcsec]^{*}$} &
\colhead{$y_c[\arcsec]^{*}$} &
\colhead{$\phi[\degr]$} &
\colhead{$q$}}
\startdata
\multirow{2}{*}{F555W} &
Lens 1 &
$  1.27 \pm  0.20$ &
$    236 \pm     18$ &
\,\,\,\,\,$   0.21 \pm   0.05$ &
$  -0.03 \pm   0.03$ &
$    154 \pm      2$ &
$  0.626 \pm  0.012$ \\ 
 &
Lens 2 &
$  0.65 \pm  0.19$ &
$    169 \pm     25$ &
$  -0.51 \pm   0.12$ &
$  -0.33 \pm   0.05$ &
$    148 \pm      4$ &
$   0.74 \pm   0.03$ \\ 
\hline \\ [-1.8ex] 
\multirow{2}{*}{F814W} &
Lens 1 &
$  1.26 \pm  0.19$ &
$    235 \pm     18$ &
\,\,\,\,\,$   0.23 \pm   0.05$ &
$  -0.03 \pm   0.03$ &
$    154 \pm      2$ &
$  0.618 \pm  0.016$ \\ 
 &
Lens 2 &
$  0.67 \pm  0.19$ &
$    171 \pm     24$ &
$  -0.55 \pm   0.13$ &
$  -0.32 \pm   0.05$ &
$    149 \pm      5$ &
\,\,$   0.76 \pm   0.03$ 
\tablecomments{Best-estimated parameter values of the lens model. Columns (from left to right) represent the Einstein radius, lensing velocity dispersion, central $x$ and $y$ coordinates, position angle, and axis ratio of the individual SIE clump. \\
$*$: Positions relative to the center of the cutout mosaic with R.A. (J2000) and decl. (J2000) of ($10^{\rm h} 11^{\rm m} 29\fs49$, $+01\degr 43\arcmin 23\farcs25$).}
\enddata
\label{tb:lens_parameters}
\end{deluxetable*}

%% file: tb_source_parameters.tex
\begin{deluxetable*}{ccccccc}
\tabletypesize{\scriptsize}
\tablewidth{\hsize}
\tablecaption{Source Model Parameters}
\tablehead{
\colhead{Band} &
\colhead{ID} &
\colhead{$m_{\rm AB} [mag]$} &
\colhead{$\mu$} &
\colhead{$R_e [\arcsec]$} &
\colhead{$R_e [\rm pc]$} &
\colhead{$n$}}
\startdata
\multirow{3}{*}{F555W} &
S1 &
$  26.7 \pm   0.6$ &
$     19 \pm     11$ &
$  0.054 \pm  0.005$ &
$    438 \pm     38$ &
$    2.4 \pm    0.3$ \\ 
 &
S2 &
$  28.1 \pm   0.1$ &
\hspace{-0.25cm} $     19 \pm      1$ &
$  0.033 \pm  0.005$ &
$    266 \pm     37$ &
$    1.4 \pm    0.4$ \\ 
 &
S3 &
$  28.2 \pm   0.5$ &
$     43 \pm     20$ &
$  0.014 \pm  0.001$ &
\hspace{-0.25cm} $    116 \pm      6$ &
$   0.22 \pm   0.03$ \\ 
\hline \\ [-1.8ex] 
\multirow{3}{*}{F814W} &
S1 &
$  26.4 \pm   0.2$ &
\hspace{-0.25cm} $     19 \pm      4$ &
$  0.055 \pm  0.006$ &
$    445 \pm     48$ &
$    2.1 \pm    0.4$ \\ 
 &
S2 &
$  27.7 \pm   0.1$ &
\hspace{-0.25cm} $     16 \pm      1$ &
$  0.024 \pm  0.002$ &
$    192 \pm     16$ &
$    0.4 \pm    0.2$ \\ 
 &
S3 &
$  28.7 \pm   0.6$ &
\hspace{-0.25cm} $    86 \pm     47$ &
$  0.013 \pm  0.008$ &
$     109 \pm     69$ &
\,\,$   0.19 \pm   0.03$ 
\tablecomments{Best-estimated parameter values of the source model. Columns (from left to right) represent the apparent magnitude, average magnification, effective radius in arcseconds and parsecs, and S{\'e}rsic index of the individual source component.}
\enddata
\label{tb:source_parameters}
\end{deluxetable*}

%% file: sect_discussion.tex
First of all, we clarify that the detected spatial offsets are between the projected 
\emph{total} mass, which includes stars, dark matter, and gas,  
and {\it V+I-}band light mostly from red, old stars in the two lens galaxies. 
In the central region that gravitational lensing is sensitive to, 
dark matter should contribute a substantial fraction to the total-mass budget. 
The lens system SDSS\,J1011$+$0143 in this work is directly comparable to 
the lens samples from the SLACS Survey \citep{SLACSV} and 
the SLACS for the Masses (S4TM) Survey \citep{Shu15} 
in terms of galaxy properties 
as they are all ETGs at similar redshifts selected from the SDSS database 
with the same selection technique. 
Studies on the SLACS and S4TM samples have shown that the projected 
dark matter fraction within the Einstein radius is $\sim 30-60\%$ 
\citep{SLACSIII, SLACSVII, SLACSX, Shu15}. 
Considering the facts that the dark matter fraction increases with 
galaxy mass/velocity dispersion and the velocity dispersions of 
the two lens galaxies in this work are at intermediate positions as compared to 
the SLACS and S4TM samples, 
we thus estimate the projected dark matter fraction of SDSS\,J1011$+$0143 
within the region enclosed by the quadruple lensed images to be 
at a level of $\sim 45\%$. 
Although the gas fraction cannot be determined with the current data, 
the detected [O\textsc{ii}] emission in the spectrum of SDSS\,J1011$+$0143 
\citep[Figure 1 in][]{Bolton06b} indicates a considerable amount of cold gas 
in this galaxy pair.

We argue that the detected mass/light offsets are real and 
not artificial owing to the limitations of our lens model. 
In this work, we use a relatively simple mass model with two SIE clumps 
for the total mass distribution. 
Suppose that the total mass distribution instead follows 
the light distribution and can be divided into four components, each of which 
is further separated into a stellar part and a dark matter part. 
Then the mass peaks of our adopted two-component model should have lain between 
light peaks 1.1 and 1.2 and between 2.1 and 2.2. 
This is obviously not the case as implied by Figure~\ref{fig:kappa}. 
We have considered a two-SIE plus external shear model but find that the external 
shear is consistent with zero and the mass/light offsets still persist. 
We also investigate the significance of the detected mass/light offsets by 
relaxing the assumption of an isothermal mass profile. 
In particular, we model the mass distributions of the two lens galaxies as two 
singular power-law ellipsoid (SPLE) clumps with the power-law index 
as an extra free parameter. We find that the two-SPLE lens model yields 
almost identical results with slightly better $\chi^2$ value (only by a few). 
To further justify that, 
we test two-SPLE lens models with mass centroids manually bound to 
the light centroids with all four possible combinations 
(note that the light centroids are still free parameters). 
These ``mass/light bound'' models all yield poorer fits with $\Delta \chi^2$ 
of $\sim 100-1200$ compared to the reported best-fit model. 
More importantly, the model with mass centroids bound to light components 
$1.2$ and $2.2$ has the smallest $\Delta \chi^2$ among all the 
``mass/light bound'' models. It again supports our finding that 
the true mass centroids are closer to the faint and more extended light peaks 
$1.2$ and $2.2$. 

The detection of significant mass/light offsets in the galaxy pair 
SDSS\,J1011$+$0143, although unexpected, is not a surprising result. 
Several lines of evidence, including the distorted morphological shapes, 
small projected separation, and presumably similar redshifts, suggest that 
the two lens galaxies have experienced/are still experiencing 
intense interactions. Stars, gas, and dark matter react 
differently during galaxy interactions. 
All materials including stars, gas, and dark matter in the interacting galaxies  
will experience dynamical friction and tidal stripping, which slow down and 
reshape the galaxies. 
Stars are effectively collisionless even under violent galaxy interactions. 
Interstellar gas is further subject to ram pressure, which exerts a drag force 
and is therefore lagged behind stars. The ``Bullet Cluster'' (1E 0657-558) is 
one of the most famous observational examples of such lagging 
\citep[e.g.,][]{Markevitch02, Clowe04, Markevitch04, Randall08}. 
Star formation activities will be triggered within the lagged and compressed 
gaseous regions. This is supported by the detected [O\textsc{ii}] emission in the 
spectra of the lens galaxies \citep[Figure 1 in][]{Bolton06b}, 
although the exact locations of star-forming regions 
are still unknown based on the current data. 

The behavior of dark matter, however, is uncertain as its dynamical nature 
is still unsettled. 
Recently, several sets of observations of the galaxy cluster Abell 3827 have 
converged to the conclusion that one of the central galaxies in 
Abell 3827 exhibits a mass/light offset of $1.62^{+0.47}_{-0.49}$ kpc 
, which was the first detection of a significant mass/light offset 
on galaxy-cluster scales \citep{Williams11, Mohammed14, Massey15}. 
The interpretation suggested by these works is that 
dark matter particles have a nonzero 
self-interaction cross section so that the collisions induce an extra 
drag force only on dark matter particles, which then get lagged behind stars. 
A constraint on the dark matter self-interaction cross section can thus be 
obtained from the amount of the offset, although the exact number is still 
subject to discussion \citep{Kahlhoefer15}. 
Being the first detection on galaxy scales, the mass/light offsets in 
SDSS\,J1011$+$0143 can also be attributed to the nonzero self-interaction 
cross-section argument.

It is worth mentioning that several studies have pointed out that 
dynamical friction could also in principle induce a similar offset 
between the spatially more extended dark matter and the more 
centrally concentrated and therefore more dynamical-friction-resistant stars 
in galaxies undergoing merging \citep{Kahlhoefer14, Mohammed14, Massey15}. 
Nevertheless, no matter which mechanisms produce the offsets, we suggest that 
interactions between the two lens galaxies are the underlying origin. 
Indeed, extensive studies of strong-lensing events in field galaxies and 
relaxed galaxy groups have shown that in general light traces mass well 
within uncertainties in terms of the coincidence of centroids 
\citep{Treu04, SLACSIII, SLACSIV, SLACSV, Dye08, Belokurov08, Limousin09, 
Suyu09, Jones10, Grillo11, Brownstein12, Oguri12, Grillo13, 
Sonnenfeld13, Grillo14, Foex14, Newman15, Shu15}. Furthermore, 
the smaller offsets in the more massive galaxy and opposite direction between 
offsets associated with individual lens galaxies in SDSS\,J1011$+$0143 cannot 
be properly explained without considering the relative motion of 
the two lens galaxies. 

Additionally, if the detected mass/light offsets are a result of 
dark matter lagging behind stars, the actual offsets between 
dark matter and stars should be larger than the reported values 
$\Delta_{1.1}$, $\Delta_{1.2}$, $\Delta_{2.1}$, and $\Delta_{2.2}$, 
which are the offsets between the \emph{total} mass and stars. 
Considering that the estimated dark matter fraction of SDSS\,J1011$+$0143 
is about $50\%$, we expect that the offsets between dark matter and stars 
are approximately twice those reported numbers.

%% file: sect_conclusion.tex
In this paper, we build detailed lens models based on high-resolution 
\textsl{HST} F555W- and F814W-band imaging data of the strong gravitational lens 
system SDSS\,J1011$+$0143, which comprises a pair of interacting ETGs at redshifts 
of $\sim 0.331$ acting as the lens and an LAE at $z_{\rm source}=2.701$ 
as the background source. The foreground-light distribution is modeled as 
four distinct S{\'e}rsic components to substantially capture the distorted 
morphological shapes due to galaxy interactions. 
To model the total-mass distribution, we 
adopt a two-SIE clump model. With three S{\'e}rsic source components, 
we obtain a satisfactory fit to the observations. 
The {\it V-} and {\it I-}band AB apparent magnitudes of the primary lens 
are $19.54 \pm 0.11$ and $17.69 \pm 0.11$, and $22.14 \pm 0.11$ and 
$20.38 \pm 0.11$ for the secondary lens. 
The best-estimated mass- and source-model parameters are in excellent agreement 
between the two filter bands. In the F555W band, the Einstein radius of 
the more massive primary lens is $1\farcs27 \pm 0\farcs20$, and 
$0\farcs65 \pm 0\farcs19$ for the relatively less massive secondary lens. 
The three highly magnified ($\times 19$ or more) source components correspond to  
intrinsically faint and compact star-forming knots in the LAE 
with {\it V-}band apparent magnitudes from 28.2 to 26.7 and 
effective radii from 116 pc to 438 pc. 
It demonstrates the capability of strong gravitational lens systems as a natural 
cosmic telescope in resolving intrinsically faint and compact objects at 
large distances.

When comparing the projected total-mass peaks inferred from lens models 
to the starlight-distribution peaks, we find significant 
spatial offsets between the mass and light of both lens galaxies. 
The offsets in the primary lens point, to the zeroth order, 
in the opposite direction of those in the secondary lens. 
The largest offset is seen in the secondary lens, with a value of 
$0\farcs36\pm0\farcs05\pm0\farcs07$ (in the F555W band) or equivalently 
$1.72\pm0.24\pm0.34$ kpc.

Such large mass/light offsets are not seen in field lens galaxies and relaxed 
galaxy groups studied in other works \citep{Treu04, SLACSIII, SLACSIV, 
SLACSV, Dye08, Belokurov08, Limousin09, Suyu09, Jones10, Grillo11, Brownstein12, 
Oguri12, Grillo13, Sonnenfeld13, Foex14, Grillo14, Newman15, Shu15}. 
We thus believe that the detected mass/light offsets in SDSS\,J1011$+$0143 
are related to the interactions between the two lens galaxies. 
Indeed, similar mass/light offsets are reported in galaxy clusters where 
galaxy interactions are even more frequent and violent 
\citep[e.g.,][]{Williams11, Mohammed14, Harvey15, Massey15}. 
These works propose that the mass/light offsets in galaxy clusters 
are caused by the separations between stars and dark matter, 
which is actually self-interacting instead of collisionless. 
They further provide estimations of the dark matter self-interacting cross section. 
However, estimation strategies are still subject to discussion 
\citep[e.g.,][]{Williams11, Harvey14, Kahlhoefer15, Massey15}. 
Although the self-interacting dark matter hypothesis would have 
a great impact on our understanding of dark matter if it were confirmed, 
other alternative explanations do exist.
For instance, several studies point out that 
dynamical friction under certain circumstances can produce 
similar offsets \citep{Kahlhoefer14, Mohammed14, Massey15}. 

To fully understand the detected mass/light offsets in SDSS\,J1011$+$0143, 
further observations and numerical simulations need to be carried out. 
The currently available \textsl{HST} F555W- and F814W-band photometric data 
are primarily sensitive to old stars in the lens galaxies, but not 
the recently born young stars and star-forming clouds, 
which should account for a substantial fraction in 
the mass budget in merging galaxies. 
Follow-up observations in the X-ray and radio can be helpful in 
determining the distribution of gas and provide a complementary view 
of this merging galaxy pair. 
So far spectroscopic data for SDSS\,J1011$+$0143 are limited to 
an SDSS observation covering the entire central $1\farcs5$-radius region 
and a Keck long-slit observation across the minor axis of the unresolved system. 
Spatially resolved (integral-field) spectra for this system would provide 
two-dimensional kinematics measurements and a lensing-independent probe of 
the total mass distribution. 
Detailed hydrodynamical simulations on galaxy-galaxy interactions, 
similar to what has been recently done by \citet{Schaller15} but with 
self-interacting dark matter considered, could quantify the relative effects of 
different mechanisms in producing the mass/light offsets observed in this system. 
Lastly, there is a significant handful of unmodeled multicomponent 
strong gravitational lenses in the SLACS and S4TM samples that 
would be a good next step for working on this topic further.